\documentclass[apjl]{emulateapj}

\shorttitle{Spitzer Observations of the Cosmic Eye}
\shortauthors{Siana et al.}

\slugcomment{Accepted for publication in ApJ}

\begin{document}

\title{Detection of PAH and Far-Infrared Emission from the Cosmic Eye:\\ Probing the Dust and Star Formation of Lyman Break Galaxies}

\author{\sc B. Siana\altaffilmark{1}, Ian Smail\altaffilmark{2}, A. M. Swinbank\altaffilmark{2}, J. Richard\altaffilmark{2}, H. I. Teplitz\altaffilmark{3}, K. E. K. Coppin\altaffilmark{2}, R. S. Ellis\altaffilmark{1}, D. P. Stark\altaffilmark{4}, J.-P. Kneib\altaffilmark{5}, A. C. Edge\altaffilmark{2}}
\altaffiltext{1}{California Institute of Technology, MS 105-24, Pasadena, CA 91125}

\altaffiltext{2}{Institute for Computational Cosmology, Durham University, South Road, Durham, DH1 3LE, UK}

\altaffiltext{3}{Spitzer Science Center, California Institute of Technology, 220-6, Pasadena, CA 91125}

\altaffiltext{4}{Institute of Astronomy, University of Cambridge, Madingley Road, Cambridge CB3 0HA}

\altaffiltext{5}{Laboratoire d'Astrophysique de Marseille, Traverse du Siphon-B.P8 13376, Marseille Cedec 12, France}

\begin{abstract}
We report the results of a $Spitzer$ infrared study of the Cosmic Eye, a strongly lensed, $L^*_{UV}$ Lyman Break Galaxy (LBG) at $z=3.074$.  We obtained $Spitzer$ IRS spectroscopy as well as MIPS 24 and 70 $\mu$m photometry.  The Eye is detected with high significance at both 24 and 70 $\mu$m and, when including a flux limit at 3.5 mm, we estimate an infrared luminosity of $L_{IR} =  8.3^{+4.7}_{-4.4}\times10^{11}$ L$_{\odot}$ assuming a magnification of 28$\pm3$.  This $L_{IR}$ is eight times lower than that predicted from the rest-frame UV properties assuming a Calzetti reddening law.  This has also been observed in other young LBGs, and indicates that the dust reddening law may be steeper in these galaxies.  The mid-IR spectrum shows strong PAH emission at 6.2 and 7.7 $\mu$m, with equivalent widths near the maximum values observed in star-forming galaxies at any redshift.  The $L_{PAH}$-to-$L_{IR}$ ratio lies close to the relation measured in local starbursts.  Therefore,  $L_{PAH}$ or $L_{MIR}$ may be used to estimate $L_{IR}$ and thus, star formation rate, of LBGs, whose fluxes at longer wavelengths are typically below current confusion limits.  We also report the highest redshift detection of the 3.3 $\mu$m PAH emission feature.  The PAH ratio, $L_{6.2}/L_{3.3}=5.1 \pm 2.7$, and the PAH-to-L$_{IR}$ ratio, $L_{3.3}/L_{IR} = 8.5\pm4.7\times10^{-4}$, are both in agreement with measurements in local starbursts and ULIRGs, suggesting that this line may serve as a good proxy for $L_{PAH}$ or $L_{IR}$ at $z>3$ with the {\it James Webb Space Telescope}.  
\end{abstract}


\section{Introduction}

Star-forming galaxies at high-redshift are often found by identifying a break in their UV continuum arising from both an intrinsic Lyman Break in their SEDs and the high opacity of the Ly$\alpha$ forest below rest-frame $1216$ \AA\ \citep{steidel96}.  Recent studies estimate that more than 25\% of all present-day stellar mass was created in these Lyman Break Galaxies (LBGs) at $z>2$ \citep{reddy09}.  Their star formation rates (SFRs) typically have to be determined based on rest-frame UV properties alone.  This involves using the UV spectral slope to determine the amount of ultraviolet extinction in order to derive the intrinsic UV luminosity and SFR.  However, many uncertainties exist in this procedure.  Firstly, there are degeneracies between age and dust reddening on the UV spectral slope.  Secondly, it is not clear that the obscuration law typically used to ``unredden'' the spectra is appropriate for LBGs.  Furthermore, at least in the local Universe, the most luminous starbursts contain individual star-forming regions that are so dusty that they effectively emit no UV light at all \citep{goldader02}, so the IR and UV properties of these systems are uncorrelated.  

Though it is more difficult to detect LBGs at other wavelengths, several studies have attempted to verify the UV-derived SFRs by comparing with other star formation diagnostics.  With near-IR spectroscopy, H$\alpha$ fluxes have been measured for $>100$ LBGs, and give comparable SFRs to the UV-derived SFRs \citep{erb06}.  However, these H$\alpha$ studies indicate that dust geometries may be different in LBGs than in local starbursts because the ionized gas does not seem to be more obscured than the stellar continuum, as is seen in local starbursts \citep{calzetti94}.  X-ray stacking of large numbers of LBGs suggests comparable SFRs as the UV determinations \citep{reddy04}, but radio continuum stacking analyses have given mixed results \citep{reddy04, carilli08}.  No individual detections of $L^{*}$ LBGs (without AGN) has been detected in the X-ray or radio.  

Ultimately, the best indicator of the star formation rate is an accurate determination of the bolometric luminosity.  Because most of the UV light (70-90\%) is absorbed by dust \citep{adelberger00, reddy06}, the majority of the starburst's luminosity is emitted thermally at infrared wavelengths.  With current technology, it is difficult to determine the IR luminosity as the SED can not be measured at multiple wavelengths.  Typical LBGs are below the confusion limit of existing submm telescopes \citep[$f_{850}<2$ mJy,][]{chapman00}, and the $Spitzer$ Space Telescope at 70 and 160 $\mu$m.  $Spitzer$ can only detect LBGs at 24 $\mu$m (rest-frame 6-8 $\mu$m) and then only $L>L^*$ LBGs are detected in the deepest images.  \citet{reddy06} conducted a study of 24 $\mu$m detected LBGs (and stacks of non-detections) and concluded that that the average UV-derived SFRs are reliable.  However, the MIPS 24 $\mu$m band only detects a few percent of the total infrared luminosity and the bolometric corrections required to convert from the mid-IR flux to $L_{IR}$ are large and highly uncertain, as the IR SEDs of LBGs have not been measured.  Therefore, in addition to determining the validity of UV-derived SFRs, it would be useful to determine if the $L_{MIR}$-to-$L_{IR}$ conversions measured locally are valid in LBGs.

A few high redshift LBGs have been found that are gravitationally lensed by foreground clusters or individual massive galaxies.  Their high magnifications (factors of 10-30) mean that their IR fluxes are above the current far-IR confusion limits and, in addition,  mid-IR spectroscopy can be performed.  These lensed LBGs can therefore be studied in the IR to better determine their star formation rates and test whether star formation and dust extinction diagnostics measured in local starbursts are valid in LBGs.  

The first detailed IR investigation of such a highly magnified LBG (MS1512-cB58) shows that the PAH strengths and the shape of the IR SED are similar to starbursts of comparable luminosity in the local universe \citep{siana08b}.  However, the IR luminosity is significantly lower than expected given the large dust extinction implied by cB58's red UV spectral slope, suggesting that the assumed dust extinction law \citep{calzetti00} may not be valid for this galaxy.  If this were true of other LBGs, it would suggest that the claimed estimates of their contribution to the star formation rate density in the early universe and, consequently, the time-integrated stellar mass density are too high.  Of course, this is only one galaxy and there is quite a large dispersion measured in the UV-IR properties of local starbursts, so IR studies of more LBGs are required.  Furthermore, cB58 is not a typical LBG in that it appears to be far younger than most LBGs ($t_{age}<30$ Myr), has a very red UV spectral slope, and displays stronger than average interstellar absorption lines.

Several other highly magnified LBGs have recently been found \citep{allam07,smail07, belokurov07,lin09}, with properties that span a broad range in parameter space occupied by LBGs (UV spectral slope, luminosity, inferred age, etc.).  Detailed IR investigations of this entire sample can determine the UV/IR properties of typical high redshift starbursts.  In this paper, we report results of a $Spitzer$ IR study of the Cosmic Eye \citep{smail07} and compare with both the cB58 findings and relations measured in local starbursts.  

The Cosmic Eye is an LBG at $z=3.074$ \citep{smail07} lensed by a massive foreground galaxy at $z=0.73$, with a total magnification $\mu = 28 \pm 3$ \citep{dye07}.  After modelling the foreground lens, reconstruction of the source image reveals the galaxy to be comprised of two UV components: a bright red and a fainter blue region.   Like cB58, the combined component photometry shows a UV slope that is redder than typical LBGs (Richard et al., in prep).  Keck integral field spectroscopy has revealed that the two UV-luminous components are part of a well ordered, rotating disk \citep{stark08}.  Detection of CO(3-2) emission indicates a large molecular gas reservoir ($\sim2\times10^9$ M$_{\odot}$) that is likely located in the fainter of the two UV components \citep{coppin07}.

We use a $\Lambda$CDM cosmology with $\Omega_m = 0.3$, $\Omega_{\Lambda}=0.7$, and $H_0 = 70$ km s$^{-1}$ Mpc$^{-1}$.  All intrinsic luminosities and star-formation rates are corrected assuming a lensing magnification, $\mu = 28\pm3$ \citep{dye07}.

\section{Observations}

$Spitzer$ IRAC and MIPS 24 $\mu$m obsevations were taken as part of Director's Discretionary Time in 2006 November/December and are detailed in \citet{coppin07}.  Additional $Spitzer$ IRS and MIPS observations were granted under Program ID 40817.  IRS Short-Low first order (7.4--14.5 $\mu$m) and Long-Low first order (19.5-38.0 $\mu$m) observations were obtained 05 December 2007.  The spectra were taken in mapping mode, placing the galaxy at five different positions along the slits.  The Short-Low (Long-Low) exposure times were 60 (120) seconds, with 50 (240) total exposures for a total of 3 (28.8) ks integration.  The IRS data reduction was performed as specified in \citet{teplitz07}.  First, we remove latent charge by fitting the slope of the increase in background with time, and subtracting this background row by row.  Second, ``rogue'' pixels were masked using the IRSCLEAN program provided by the SSC.  Finally, the observations at other map positions were used to determine the sky, which was then subtracted.  The individual frames were co-added to produce 2D spectra at each map position.  One-dimensional spectra were optimally extracted at each map position using the SPICE software provided by the SSC.  

The MIPS 70 $\mu$m observations were taken on 28-29 November 2007.  1080 exposures at nine dithered positions were taken for a total of 10.8 ks integration time.  The MIPS 70 $\mu$m data were reduced using the Germanium Reprocessing Tools (GeRT), following the techniques optimized for deep photometry data given by \citet{frayer06}.  The images were then mosaiced with MOPEX \citep{makovoz05b} and extracted with APEX \citep{makovoz05a}.

The IRS Peak-Up Imaging 16 $\mu$m observations were taken on 06 December 2007.  Twenty dithered exposures of 30 seconds were taken for a total of 600 s integration.  A median sky was created and subtracted from each BCD after scaling to the mode.  The BCDs were combined using MOPEX, using both temporal and spatial outlier rejection (Mosaic Outlier and Dual Outlier).  Interpolation was performed using the drizzle algorithm with {\sc Driz\_Frac}$=0.8$ and an output pixel size of 0.9$''$ (half of the native PUI plate scale).  We used APEX for source extraction, with a custom PRF made with the same drizzle parameters.

\begin{deluxetable}{lrr}
\tablecaption{Cosmic Eye Photometry}
\tablehead{\colhead{Band} & \colhead{Flux Density} & \colhead{Error\tablenotemark{a}}\\
	       \colhead{[$\mu$m]} & \colhead{[$\mu$Jy]} & \colhead{[$\mu$Jy]}}
\startdata
16	&	90 & 18 \\
24	&	281\tablenotemark{b} & 65\tablenotemark{b} \\
70	&	4100 & 1300 \\
\enddata
\tablenotetext{a}{Errors are $1\sigma$}
\tablenotetext{b}{From \citet{coppin07}}
\label{tab:photom}
\end{deluxetable}

\section{Results}

\subsection{Infrared SED and $L_{IR}$}
\label{ir_sed}
The 16, 24, and 70 $\mu$m photometry all yield greater than $5\sigma$ detections (see Table \ref{tab:photom}).  We use these fluxes, combined with a 3.5 mm flux limit from \citet{coppin07} to fit the shape of the IR spectral energy distribution (SED) and determine the infrared luminosity.  The diameter of the Cosmic Eye is $\sim 2''$ and is therefore unresolved in any of the IRS and MIPS photometry (the 16, 24, and 70 $\mu$m PSF FWHMs are 4$''$, 6$''$, and 18$''$, respectively).  Therefore, the foreground galaxy at $z=0.73$ may be contaminating the mid-IR photometry.  Here we attempt to determine the magnitude of this foreground contamination.  

\begin{figure}
\epsscale{1.0}
\plotone{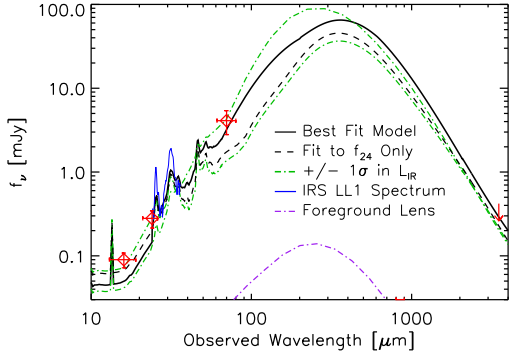}
\caption{The observed IR SED of the Cosmic Eye.  The best-fit SED from \citet{chary01} to the 24, 70 $\mu$m and 3.5 mm fluxes yields $L_{IR} = 8.3\times10^{11}$ L$_{\odot}$.  Also plotted is the SED derived from the 24 $\mu$m flux alone, the SEDs giving the $\pm1\sigma$ deviations in $L_{IR}$, and the estimated IR SED of the foreground lens (based on the [O{\sc ii}] flux).  \label{fig:ir_sed}}
\end{figure}

In the optical spectrum obtained in \citet{smail07}, the foreground galaxy has an [O{\sc ii}] 3727\AA\ emission line flux $f_{[O{\sc II}]} \sim 1.5\times10^{-17}$ erg s$^{-1}$ cm$^{-2}$, or $L_{[O{\sc II}]} = 3.6\times10^{40}$ erg s$^{-1}$.  Using the \citet{kennicutt98} conversion to star-formation rate, we get SFR$_{fg}$([O{\sc ii}])$ \sim 0.5$ M$_{\odot}$ yr$^{-1}$.  This assumes an extinction of $A_{V}=1$.  Converting this SFR to an $L_{IR}$ using \citet{kennicutt98}, we get an $L_{IR} = 2.3\times10^{9}$ L$_{\odot}$.  If we choose an IR template typical of galaxies with this SFR \citep{chary01}, the expected fluxes of the lens at 16, 24, and 70 $\mu$m are more than an order of magnitude lower than the observed fluxes (see Figure \ref{fig:ir_sed}).  Of course SFRs (and $L_{IR}$) based upon [O{\sc ii}] 3727\AA\ flux alone are quite uncertain due to unknown extinction and metallicity, but it seems very unlikely that the extinction is an order of magnitude higher than $A_{V}\sim1$ in such an evolved massive galaxy.  

As an additional check of foreground contamination we also use the observed 16 $\mu$m flux to estimate the maximum IR contamination.  The 16 $\mu$m band samples the minimum of the Cosmic Eye SED at $\lambda_{rest}= 4$ $\mu$m where both the stellar and dust SEDs are faint. However, the 16 $\mu$m band samples the possibly significant PAH emission from the foreground galaxy.  By scaling the stellar SED to the IRAC bands we find that the majority ($>65$\%) of the measured $f(16\mu$m) is from the stars in the Cosmic Eye.  This gives a conservative upper limit $f_{lens}(16\mu$m$)<0.03$ mJy from the lens alone.  This is in good agreement with the SED assumed when deriving the SFR in the lens from the [O{\sc ii}] 3727\AA\ flux.  Therefore, estimates from both [O{\sc  ii}] 3727\AA\ flux and the 16 $\mu$m flux show that the lens' contribution to flux at 24 and 70 $\mu$m is negligible.  

Following the discussion above, we assume that the measured Spitzer fluxes are dominated by the Cosmic Eye.  We fit IR SED templates from \citet{chary01} to the 24 and 70 $\mu$m photometry, as well as the 3.5 mm $1\sigma$ limit ($\lambda_{rest}\sim 870 \mu$m) from the CO observations \citep[$f_{3.5mm}<0.14$ mJy,][]{coppin07}.   The best-fit template is a warm IR SED with a magnification corrected $L_{IR} = 8.3\times10^{11}$ L$_{\odot}$ (See Figure \ref{fig:ir_sed}).  Other SED shapes are allowed that give a $1\sigma$ range of $L_{IR}=3.9$--$13\times10^{11} L_{\odot}$.  Using the conversion of \citet{kennicutt98} we derive a $SFR_{IR} = 140 \pm 80$ M$_{\odot}$ yr$^{-1}$.  If we select a template appropriate for the measured mid-IR luminosity, $L_{MIR}$, based on $f(24\mu$m) alone (as is often done at high redshift),  we derive an infrared luminosity nearly a factor of two smaller (see dashed line in Figure \ref{fig:ir_sed}, $L_{IR} = 4.8\times10^{11}$ L$_{\odot}$). It's important to note that none of the templates gives an $L_{IR}$ larger than $1.3\times10^{12}$ L$_{\odot}$.  This is because no cold or warm dust can be added without further violating the measurements at 3.5 mm or 70 $\mu$m, respectively.   Because these fluxes may also have some foreground lens contamination, we take $L_{IR}<1.3\times10^{12}$ L$_{\odot}$ as a conservative upper limit.  

\subsection{Infrared Spectrum and PAH Luminosities}

\begin{figure}
\epsscale{1.0}
\plotone{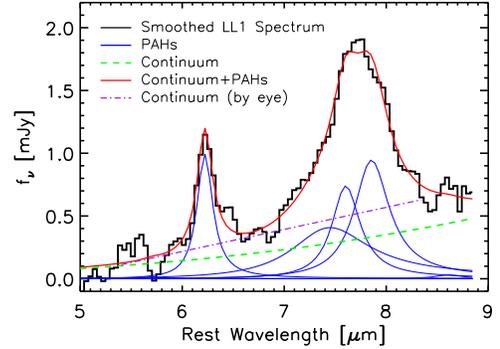}
\caption{The smoothed IRS Long-Low spectrum of the Cosmic Eye.  The simultaneous fit of the PAH features and continuum is also plotted.  The dot-dashed line is the continuum assumed when computing PAH fluxes to compare with \citet{pope08}. \label{fig:pahfit}}
\end{figure}

The IRS Long-Low spectrum is plotted in Figure \ref{fig:pahfit}.  We see prominent PAH emission at rest-frame 6.2 and 7.7 $\mu$m.  Unfortunately, the 8.6 $\mu$m feature lies close to the noisy end of the spectrum so its amplitude is uncertain.  The PAH strengths were measured by simultaneously fitting Drude profiles with centers at the systemic redshift of the galaxy and widths defined by \citet{draine07a}, as well as a power-law continuum with a slope that is allowed to vary. The best fit components are plotted in Figure \ref{fig:pahfit} and the derived line fluxes are listed in Table \ref{tab:lines}.  Some authors have determined PAH luminosities and equivalent widths by simply assigning a continuum value based on the fluxes immediately longward and shortward of the features.  Fluxes measured in this way are typically lower by up to a factor of two.  These fluxes are also listed in Table \ref{tab:lines}.  

\begin{figure}
\epsscale{1.0}
\plotone{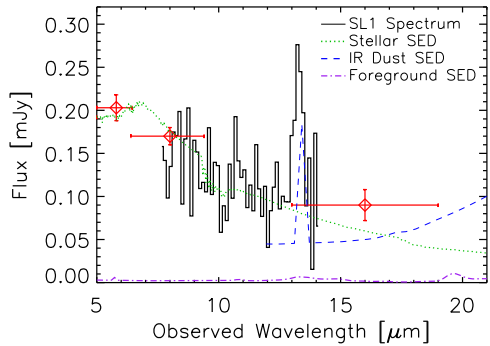}
\caption{The IRS Short-Low spectrum of the Cosmic Eye.  The feature at 13.3 $\mu$m is either 3.3 $\mu$m PAH emission at $z=3.07$ or 7.7 $\mu$m PAH emission at $z=0.73$.  A stellar SED from a 100 Myr old (constant star formation) starburst is scaled to the IRAC bands.  The SL spectrum was scaled up by 30\% to match this SED.  The best-fit \citet{chary01} SED fit to the far-IR data is shown as well as a star-forming SED at $z=0.73$ with an $L_{IR}=2.3\times10^9 L_{\odot}$ estimated from the foreground lens' [O{\sc ii}] emission line.  The 3.3 $\mu$m PAH at $z=3.074$ is expected to be much stronger than the 7.7 $\mu$m PAH at $z=0.73$.  \label{fig:sl1}}
\end{figure}

The IRS Short-Low spectrum is plotted in Fig \ref{fig:sl1}.  The SL1 spectrum covers 7.5-14 $\mu$m and covers the Pa$\alpha$ ($\lambda_{rest}=1.875$ $\mu$m) and 3.3 $\mu$m PAH emission lines at $z=3.074$.  The Pa$\alpha$ line is not seen but this is not surprising as an intrinsic SFR $\sim 140$ M$_{\odot}$ yr$^{-1}$ gives an expected flux $f(Pa\alpha) = 2.3\times10^{42}$ erg s$^{-1}$ (assuming a 28$\times$ magnification) if the line is free of dust extinction.  Therefore, we would only expect to detect it at less than $1.5\sigma$.  Here we have assumed case B recombination and convert from H$\alpha$-to-SFR conversion of \citet{kennicutt98} using P$\alpha$/H$\alpha$ = 0.128 \citep{hummer87}.  

\begin{figure}
\epsscale{1.0}
\plotone{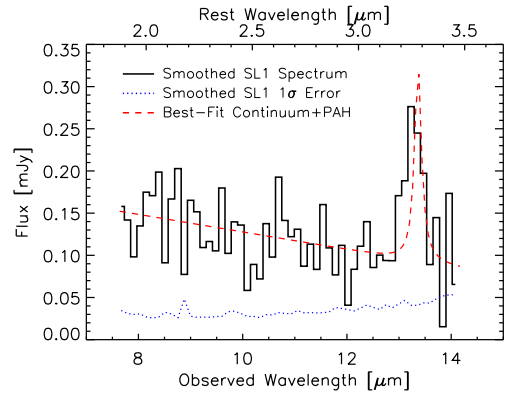}
\caption{The Short-Low spectrum after binning by 2 pixels as well as the binned errors.  The combined fit (to the unbinned data) of the 3.28 $\mu$m drude profile and a linear continuum is also plotted.  The 3.3 $\mu$m PAH feature is detected at 5.3$\sigma$. \label{fig:sl1_pah33}}
\end{figure}

An emission line is seen at $\lambda_{obs} = 13.3$ $\mu$m.  This can either be the 3.3 $\mu$m PAH feature at $z=3.074$ or the 7.7 $\mu$m PAH line at the redshift of the foreground lens ($z=0.73$).  We believe that the feature is unlikely to be from the foreground lens.  First, the flux falls off too quickly at $\lambda_{obs} > 13.5$ $\mu$m, inconsistent with the broad 7.7 $\mu$m feature.  Second, as seen in Figure \ref{fig:sl1}, nearly all of the 16 $\mu$m flux can be explained by the stellar and dust emission from the Eye, without a significant contribution from PAHs from the foreground lens.  So it is unlikely that the PAHs of the foreground lens are any stronger than the estimated foreground SED plotted in Figure \ref{fig:sl1} (see $\S$ \ref{ir_sed}).  In Figure \ref{fig:sl1_pah33} we show the binned (by 2 pixels) Short-Low spectrum with the best fit (linear) continuum plus the 3.3 $\mu$m PAH profile and list the flux in Table \ref{tab:lines}.  The 3.3 $\mu$m PAH feature is significant at 5.3$\sigma$. 

\begin{deluxetable*}{lrrr}
\tablecaption{Cosmic Eye Line Fluxes}
\tablehead{\colhead{Rest Wavelength} & \colhead{Observed Flux\tablenotemark{a}} & \colhead{Luminosity\tablenotemark{a}\tablenotemark{b}} & \colhead{Rest Equivalent Width} \\
 \colhead{[$\mu$m]} & \colhead{[$10^{-15}$ ergs s$^{-1}$ cm$^{2}$]} & \colhead{[$10^{42}$ ergs s$^{-1}$]} & \colhead{[$\mu$m]}}
\startdata
1.87 Pa$\alpha$   & $<$0.41				     & $<$1.2                               & $<$0.13 \\
3.3  PAH               & 0.92$\pm0.17$		     & 2.7$\pm0.5$                      & 0.13 \\
6.2  PAH	            & 5.36$\pm1.23$                     & 15.7$\pm3.6$                    & 1.7 \\
                            & (3.86)\tablenotemark{c}          & (11.3)\tablenotemark{c}        & (0.78)\tablenotemark{c}  \\
7.7  PAH               & 17.9$\pm4.48$                    & 52.4$\pm13.3$                  & 4.4 \\
                            & (13.8)\tablenotemark{c}          & (40.3)\tablenotemark{c}        & (2.05)\tablenotemark{c} \\
8.6  PAH               & 0.1$\pm2$			     & 0$\pm5$                            &
\enddata
\tablenotetext{a}{\ Errors and limits are $1\sigma$}
\tablenotetext{b}{\ Intrinsic luminosity after correction for assumed 28$\times$ magnification}
\tablenotetext{c}{\ Values in parentheses are determined assuming the purple line in Figure as continuum \ref{fig:pahfit} for continuum as in \citet{pope08}.}
\label{tab:lines}
\end{deluxetable*}

\section{Discussion}

The {\it Spitzer} data have confirmed strong PAH emission at 3.3, 6.2, and 7.7 $\mu$m.  In addition, the 24 and 70 $\mu$m photometry, in combination with the 3.5 mm flux limit give a good estimate of the shape of the IR SED and the $L_{IR}$ (and the IR-derived SFR).  In this section we use these IR characteristics, as well as ancillary rest-frame UV and optical spectra, to compare with local star-forming galaxies of comparable luminosity as well as high redshift, submm-selected ultraluminous galaxies.  

\subsection{Infrared Excess vs. UV Spectral Slope}

We would like to determine whether the UV- and IR-derived SFRs agree if we assume the same relations measured in local starbursts.  Typically, the UV spectral slope is used to determine the UV extinction, $A_{1600}$.  The intrinsic UV luminosity can then be calculated and used to determine the UV-derived star formation rate, SFR$_{UV}$.  We measure the UV spectral slope $\beta$ (where $f_{\lambda} \propto \lambda^{\beta}$) in two ways: with broadband photometry, and from the rest-frame UV spectrum itself.  Broadband photometry mimics the method used for most LBGs, especially those that are fainter and/or at higher redshifts.  It may seem better to derive the spectral slope directly from the spectrum, but the spectrum may suffer from differential atmospheric dispersion, only samples part of the total lens, and may have a small amount of contamination from the foreground lens (though this should be less than a few percent of the total flux).  First we use the high spatial resolution HST photometry of Richard et al. (2009, in prep) and subtract a fit of the foreground lens profile.  The resulting color, $V_{606}-I_{814} = 0.53\pm0.06$ (AB) gives an estimate of the spectral slope according to Eqn. 14 of \citet{meurer99}.  This equation takes into account the effect of interstellar absorption lines and the the Ly$\alpha$ forest opacity below $\lambda_{rest}<1216$ \AA\ (which affects a small fraction of the $V_{606}$ band at this redshift).  We also fit the spectral slope to the parts of the optical spectrum that are uncontaminated by absorption lines.  The spectrum is a composite of four parts of the entire ring \citep[see Figure 1 of][]{smail07}, not the entire galaxy.  Therefore small differences in spectral slopes based on the photometry and spectra are expected.  The best-fit slope from the spectrum, $\beta_{spec} = 0.01$, is redder than the slope derived from the photometry alone, $\beta_{photom}=-0.48$.  We take the average of the two values and use the difference as the $\pm1\sigma$ range.  The resulting spectral slope, $\beta = -0.24\pm0.24$, is very red for LBGs, such that its observed optical colors lie near the edge of typical Lyman Break color selection criteria.  

\citet{meurer99} determined a relation between the UV spectral slope and the UV attenuation, $A_{1600}$.  Essentially, this assumes that the starburst has an intrinsic spectral slope $\beta \sim -2.3$, and that the shape of the attenuation curve, $A_{\lambda}$, is that of the Calzetti reddening law \citep{calzetti97}.  The \citet{meurer99} relation suggests a UV attenuation, $A_{1600} = 3.95\pm0.74$ mags (the error here includes the 0.55 mag dispersion in the observed $\beta$-to-$A_{1600}$ relation as well as the error in the spectral slope measurement).  Correcting for this attenuation (and for the lens magnification), the measured UV luminosity is $L_{1500} = 6.3\times10^{30}$ erg s$^{-1}$ Hz$^{-1}$.  This translates to a star formation rate, SFR$_{UV} = 900^{+900}_{-400}$ M$_{\odot}$ yr$^{-1}$, using the conversion of \citet{kennicutt98}.  

\begin{figure}
\epsscale{1.0}
\plotone{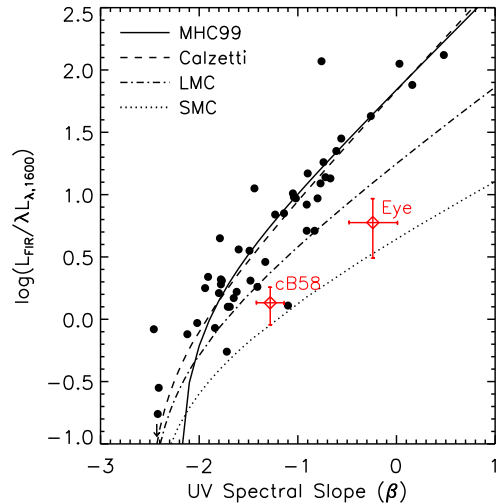}
\caption{The far-infrared (40-120 $\mu$m) to ultraviolet (1600 \AA) luminosity ratio versus the UV spectral slope, $\beta$, defined as $f_{\lambda}\propto \lambda^{\beta}$.  The best-fit relation of MHC99 and measurements of local starbursts are plotted.  We show the expected IRX-$\beta$ relation for three different reddening curves (Calzetti, LMC with out the 2175 \AA\ feature, and SMC).  The LMC and SMC extinction curves more accurately predict the observed FIR-to-UV ratios of the two lensed LBGs. \label{fig:irx_beta}}
\end{figure}

The UV-derived SFR is about six times that of the IR-derived SFR ($SFR_{IR} = 140 \pm 80$) and even the $1\sigma$ lower limit is three times higher.  The red UV spectrum suggests too much UV attenuation, which results in a predicted infrared luminosity far above the maximum allowed $L_{IR}$ from the observations.  The same phenomenon was observed with the first lensed LBG to be studied in detail in the infrared \citep[cB58,][]{siana08b} as well as in $Spitzer$ 24 $\mu$m studies of unlensed, young ($t_{age}<100$ Myr) LBGs \citep{reddy06}.  In Figure \ref{fig:irx_beta}, we plot the far-IR to UV luminosity ratio versus UV spectral slope, $\beta$, fit to a sample of local starburst galaxies \citep[IRX-$\beta$,][]{meurer99}.  This relation follows naturally assuming intrinsic spectral slopes typical of young starbursts ($-2.6<\beta<-2.0$) and a Calzetti reddening law \citep{calzetti97}.   Because LBGs are assumed to have similar intrinsic spectral slopes as these local starbursts and are assumed to be reddened by something like a Calzetti law, this relation is generally used to infer infrared luminosities and, thus, star formation rates.  In Figure \ref{fig:irx_beta}, both cB58 and the Cosmic Eye lie below this relation by at least a factor of four, beyond the typical scatter in this relation that is observed locally.

Of course, many local star-forming galaxies are known to lie off of this IRX-$\beta$ relation. If much of the star-forming regions are completely extinguished in the UV, then the observed UV spectral slope will no longer be correlated with IRX.  \citet{goldader02} showed this to be true in local ULIRGs as they all lie above the IRX-$\beta$ relation of \citet{meurer99}.

If the intrinsic spectral slope is different due to different metallicities or star formation histories, this will cause galaxies to move significantly away from the measured relation.  For example, less active star-forming galaxies such as nearby spirals are known to fall to the right of the IRX-$\beta$ relation because their relatively large amount of less massive (older) stars makes the intrinsic spectral slope redder \citep{bell02}.  

Finally, if the reddening curve is significantly different than the assumed Calzetti reddening curve, this will affect the total amount of absorbed luminosity for a given UV spectral slope.  For example, the SMC extinction curve \citep{prevot84} is much steeper than the Calzetti law and can produce red UV slopes without extinguishing as much total luminosity as with a Calzetti law.  Therefore, if the dust extinction in a particular galaxy follows an SMC curve, its IR luminosity would be far lower for a given $\beta$ and would lie below the IRX-$\beta$ relation.  

Because these LBGs are so well characterized, we can investigate whether any of their known properties might explain why they lie below the relation observed in local starbursts.  Using the $R_{23}$ index \citep{pettini04, pilyugin05}, \citet{stark08} determine a metallicity for the Cosmic Eye of $\sim0.9$ Z$_{\odot}$.  SED fits to the Cosmic Eye optical and near-IR photometry (Richard et al., in prep) give ages of 80-300 Myr depending upon the exact star formation history.  Both the metallicity and starburst age of the Cosmic Eye are similar to that of the local starburst sample and the estimated intrinsic UV spectral slope is $\beta\sim -2.4$, similar to the intrinsic slopes estimated for the local sample.  cB58 has a metallicity of about $\frac{1}{3} Z_{\odot}$ \citep{pettini00, teplitz00} and a very young starburst age \citep[$t_{age} < 30$ Myr,][]{ellingson96,siana08b}.  The \citet{bruzual07} models give an intrinsic slope of $\beta \sim -2.7$ for this metallicity and star formation history, bluer (more negative) than assumed for the local starbursts.  Therefore, cB58 should lie to the {\it left} of the local relation, not to the right.  

Given the intrinsic slopes predicted by their star formation histories and metallicities, both the Eye and cB58 should lie near (or to the left) of the local IRX-$\beta$ relation if the Calzetti reddening law  \citet{calzetti97} were valid in these LBGs.  Therefore, we conclude that the reddening law in these two LBGs must be different than that measured in the local starburst sample.  In Figure \ref{fig:irx_beta} we show three predicted IRX-$\beta$ relations assuming different reddening curves: Calzetti, LMC \citep{fitzpatrick86} with no 2175 \AA\ feature, and SMC.  A reddening law that is steeper than the Calzetti curve, like the LMC or SMC, causes UV spectral slopes to become redder without extinguishing as much total UV luminosity.  Thus, for the same observed UV spectral slope, the steeper reddening curves result in lower $L_{FIR}/L_{UV}$ than predicted by the Calzetti curve.  As seen in Figure \ref{fig:irx_beta}, both the Cosmic Eye and cB58 have $L_{FIR}/L_{UV}$ ratios indicative of steeper reddening laws. 

\citet{siana08b} conclude that the large covering fraction of outflowing, low-ionization gas (and presumably dust) seen in the rest-frame UV spectrum of cB58 is indicative of a uniform foreground sheet of dust, rather than a patchy distribution that gives rise to a Calzetti curve.   The rest-frame UV spectrum of the Cosmic Eye also exhibits opaque interstellar absorption lines indicating a similarly large covering fraction of outflowing gas.  A uniform foreground sheet of dust results in reddening laws analogous to the LMC or SMC extinction curves (assuming similar dust compositions).  As shown above, these steeper curves can explain why these LBGs fall below the local IRX-$\beta$ relation.    Therefore positions of the Cosmic Eye and cB58 on the IRX-$\beta$ diagram can be explained if much of the dust obscuration is occuring in outflowing dust with a large covering fraction. 

Alternatively, a different reddening curve could be caused by a different dust composition.  It is possible that in a young galaxy with active star formation, a larger fraction of the dust is produced by core collapse supernovae relative to dust produced by Type Ia SNe or evolved, low-mass (eg. AGB) stars.  Evidence for extinction due to dust from Type II SNe \citep{todini01} has been observed in host galaxies of a QSO and a gamma-ray burst at $z\sim6.2-6.3$ \citep{maiolino04,stratta07}, when the age of the universe is less than the time required for stars to evolve to the AGB phase (when much of the dust is deposited into the ISM).  Because the star formation activity in these two LBGs is recent ($<300$ Myr), this could imply that the fraction of dust from Type II SNe to dust from AGB stars is larger in these systems, which may affect the shapes of the reddening curves.  However, both LBGs show strong emission from PAHs, which are thought to be produced primarily in AGB stars \citep{latter91,tielens08}, so it is possible that there exists a population of less luminous, older stars that is also depositing dust into the ISM.

It is impossible to make any broad conclusions about the LBG population as a whole based on these two LBGs alone.  However, it is instructive to note that at least a subset of LBGs may not obey the typical relations assumed for all LBGs.  It appears that UV-derived SFRs of young LBGs with strong interstellar absorption features may be overestimated by a factor of $\sim4-5$.  According to SED fits to the rest-frame UV-optical photometry, $\sim$20\% of all $\sim L^*$ LBGs have starburst ages less than 100 Myr \citep{shapley01,papovich01}.  Therefore, any large (factor $\sim4$) adjustment to their derived star formation rates will significantly impact the total SFR density at high redshift.  

Of course, many high redshift galaxies also lie {\it above} the local relations such that their UV properties {\it under}predict their total SFRs \citep{reddy06}.  However, many of these galaxies (eg. submm continuum selected galaxies) are accounted for separately when determining star-formation rate densities at high redshift \citep{chapman05}.      

\subsubsection{Caveats}

There are a few complications in our analysis that arise because we are observing a lensed galaxy.  The first is that the UV-luminous portions of the eye may lie closer to the caustic than the IR-luminous regions and are therefore more highly magnified, resulting in the source lying below the local IRX-$\beta$ relation.  This is especially relevant as the Eye was selected for its bright rest-frame UV fluxes, not its IR fluxes, so we may be biased toward galaxies with high UV magnifications.  Source reconstruction of the Cosmic Eye shows that the most UV luminous region lies near the caustic and is thus highly magnified \citep{dye07}.  If the central region of the galaxy, which is further from the caustic \citep[see Figure 2 in][]{stark08}, hosts a more obscured starburst (as is commonly observed in local LIRGs), this region will not be so highly magnified, and the observed $L_{IR}$-to-$L_{UV}$ ratio will not be the same as the ratio observed without the foreground lens.  

Of course, for this phenomenon to bias our results, the small-scale star-forming regions within the LBGs must violate the local IRX-$\beta$ relationship (ie. the UV and IR fluxes are uncorrelated at sub-kpc scales).  There is no evidence that this is the case in local galaxies.  For example, individual star-forming regions in M51a appear to show similar trends at scales $<500$ pc \citep[though the trend is offset due to the presence of older stellar populations,][]{calzetti05}.  If we assume that the UV and IR emission in LBGs are also correlated on scales smaller than 500 pc, then the effect of differential magnification on our results is mitigated significantly.  

It is also possible that there is dust present around the foreground lens that is further reddening the UV spectrum of the background LBG.  This would affect the observed UV spectral slope, but would not significantly increase the observed $L_{IR}$.  This would require significant columns of dust at large radii from the center of the foreground lensing galaxies ($\sim7$ physical kpc for the lens of the Cosmic Eye).  There are many galaxies with dust at $>7$ kpc.  For example, \citet{engelbracht06} find emission from PAHs in supernovae winds at $>6$ kpc from M82.  The Sa galaxy, the Sombrero Galaxy, and the Sd galaxy, NGC 4594, both have dust at radii of $\sim 7$ kpc \citep{bendo06,bendo06b}.  Also, recent GALEX observations have found low levels of star-formation (where some dust is presumably present) in the extreme outer disks of M83 and NGC 4625  \citep[$\sim 10$ kpc,][]{thilker05,gildepaz05}.  However, these dust features and low level star-forming regions are found in large spiral disks (eg. M83) or actively star-forming galaxies (M82), and are not typically found around massive ellipticals.  Furthermore, when the dust is present, it is patchy and distributed along spiral arms and has low optical depths at the relevant wavelengths \citep[$\tau(4000$ \AA$) < 0.3$,][]{holwerda09}.  Given the high stellar mass and low star formation rate of the foreground lens of the Cosmic Eye and the fact that the color changes very little around the ring of the Eye, we expect the foreground extinction to be negligible.  

\subsection{PAH Properties}

\subsubsection{Mid-IR Spectra Comparison}
\begin{figure}
\epsscale{1.0}
\plotone{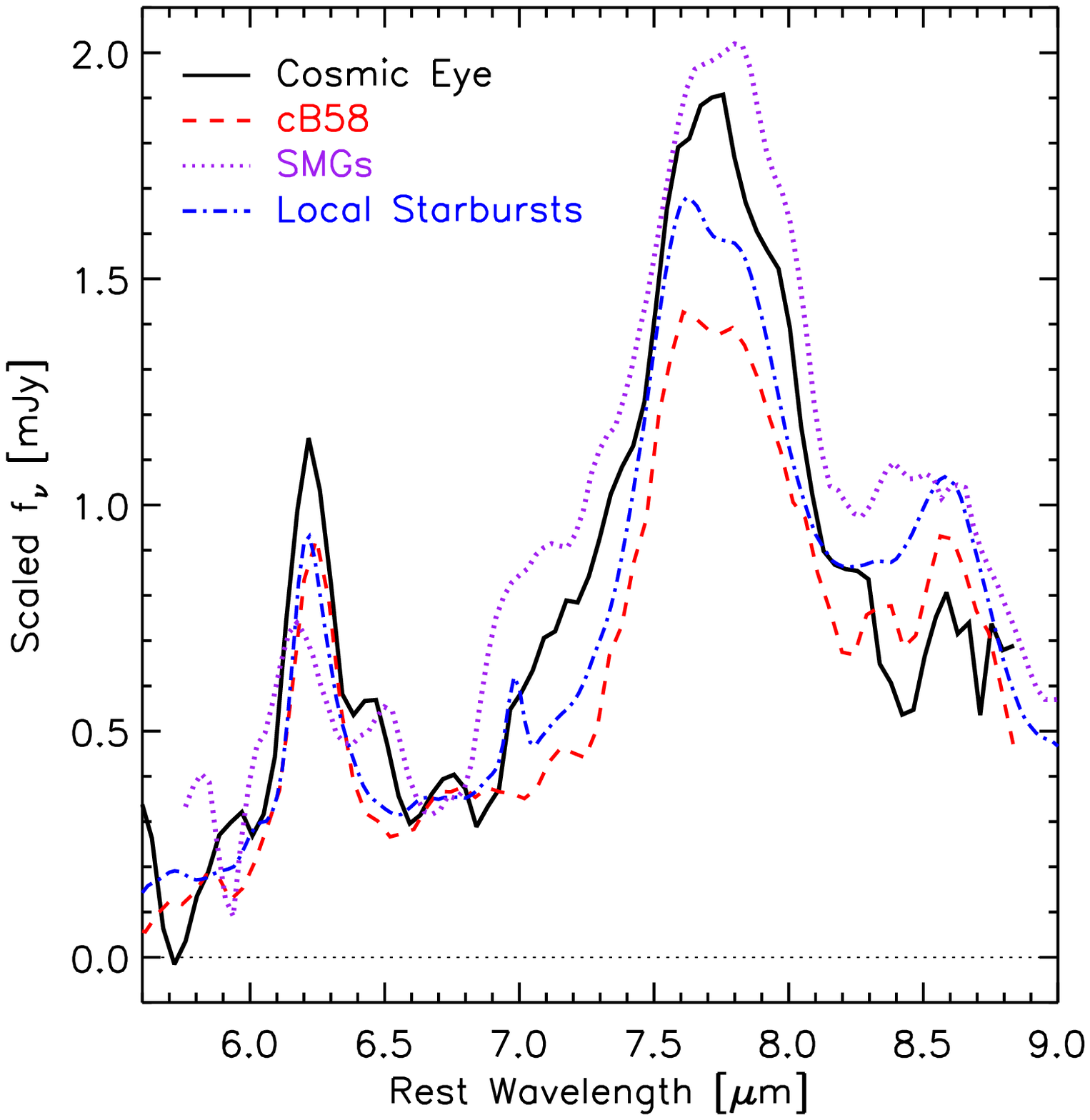}
\caption{Comparison of mid-IR spectra (normalized to the 6.7 $\mu$m continuum) of the Cosmic Eye, cB58 \citep{siana08b}, an SMG composite \citep{menendez-delmestre09}, and a local starburst composite \citep{brandl06}.  These galaxies span three orders of magnitude in luminosity in $L_{IR}$ ($10^{10}-10^{13}$ $L_{\odot}$).  The mid-IR spectra of the LBGs (Cosmic Eye and cB58) and the local starburst composite are similar.  However, the $L_{7.7}/L_{6.2}$ ratio of the SMGs is markedly higher than in the LBGs or local starbursts. \label{fig:spec_comp}}
\end{figure}

In Figure \ref{fig:spec_comp} we compare the mid-IR spectra of the Cosmic Eye with another lensed LBG \citep[cB58,][]{siana08b}, a composite of high redshift submm-selected ULIRGs \citep{menendez-delmestre09}, and a composite of low redshift, lower luminosity ($L_{IR}<10^{11}\ L_{\odot}$) starburst galaxies \citep{brandl06}.  All of the spectra exhibit high PAH equivalent widths and the PAH ratio, $L_{7.7}/L_{6.2}$, is approximately constant in all of these spectra except for the SMG composite, which has a markedly higher ratio than the others.  This PAH ratio does not change with ionization state of the PAHs, but changes dramatically with grain size distribution \citep{draine07a}.  The discrepancy may suggest that grain sizes are relatively larger in SMGs than in LBGs and local starbursts.  However, there is significant dispersion in all of the populations so a comparison of one or two LBGs to the SMG sample as a whole is not definitive.  

Because the 7.7 $\mu$m feature is confused with the 8.6 $\mu$m feature and the broad silicate absorption at 9.7 $\mu$m, we choose to use the 6.2 $\mu$m PAH EW for comparison with other works.  In Table \ref{tab:lines}, we list the derived rest-frame equivalent widths (EWs) of the PAHs using two separate methods: by simply defining a linear continuum on either side of the PAHs and by simultaneously fitting all PAHs and a power-law continuum.  The other studies to which we are comparing have derived EWs in a manner similar to the former method.  Therefore, we use the the equivalent widths listed in parantheses in Table \ref{tab:lines} for comparison.  The rest-frame equivalent width, $EW_{6.2}=0.78$ $\mu$m, is very high, near the maximum values found in local star-forming galaxies with no AGN activity \citep{brandl06,desai07,imanishi07} as well as high redshift ULIRGs with strong PAH emission \citep{sajina07,pope08}.

\subsubsection{$L_{PAH}$ vs. $L_{IR}$}
\begin{figure}
\epsscale{1.0}
\plotone{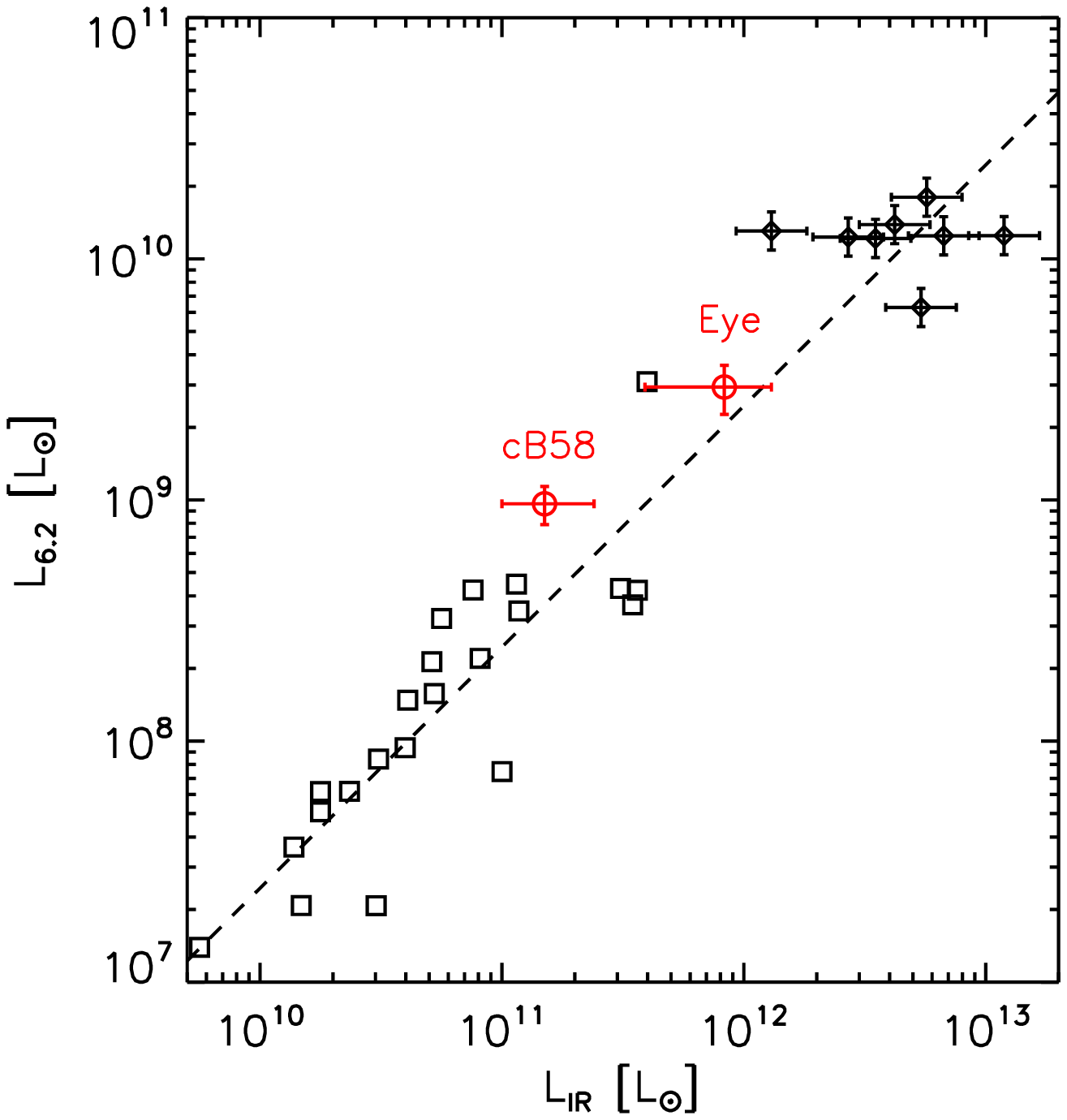}
\caption{The 6.2 $\mu$m PAH strength vs. infrared luminosity.  This figure is adapted from the top panel of Fig. 12 in \citet{pope08}.  Local starbursts (squares) are from \citet{brandl06} and high-redshift SMG (diamonds) are from \citet{pope08}.  The dashed line is the best-fit $L_{6.2}$-to-$L_{IR}$ relation of \citet{pope08} for the low redshift starbursts.  The $L_{PAH}/L_{IR}$ ratios of the two lensed LBGs are consistent with the extrapolation (to higher luminosities) of the trend observed in local starbursts.  \label{fig:l62_lir}}
\end{figure}

In Figure \ref{fig:l62_lir} we plot the 6.2 $\mu$m PAH luminosity vs. $L_{IR}$ for the Eye relative to local starbursts \citep{brandl06}, high redshift SMGs \citep{pope08} and cB58 \citep{siana08b}.  We have remeasured our PAH luminosities in a similar manner as that of \citet{pope08} by selecting the continuum level on either side of the PAH features, rather than simultaneously fitting all features and the continuum.  These PAH luminosities are also listed in Table \ref{tab:lines}.  Both cB58 and the Cosmic Eye lie above the measured relation of \citet{pope08} but certainly within the rather large scatter.  Therefore, for these two LBGs, it appears that the 24 $\mu$m flux, which is dominated by PAH emission, would give a reasonable estimate of the $L_{IR}$, lending credence to high redshift ($1.5<z<3.0$) SFR determinations based on $Spitzer$ 24 $\mu$m fluxes alone.  A similar study of lensed galaxies of somewhat higher IR luminosities at high redshift has also found that rest-frame 8 $\mu$m fluxes also reproduces the $L_{IR}$ reasonably well \citep{rigby08}.  

\subsubsection{3.3 $\mu$m PAH}
We have a $5\sigma$ detection of the 3.3 $\mu$m PAH feature.  This is the highest redshift galaxy in which this feature has been detected, and only the second \citep[the first was reported in][]{sajina09} to be detected outside of the local ($z<0.2$) universe \citep{moorwood86,imanishi00,imanishi06a,risaliti06}.  The 3.3 $\mu$m PAH feature will be particularly useful in future studies of dust and star formation as it is the only PAH emission feature accessible to the {\it James Webb Space Telescope} (JWST) at $z>3$.  Therefore, it is important to characterize how well $L_{3.3}$ scales with other PAHs and $L_{IR}$ in LBGs at $z<3$ so that JWST studies can properly interpret measurements at higher redshift.  The PAH ratio, $L_{6.2}/L_{3.3} = 5.8 \pm 1.7$ is consistent with the typical ratio measured in local ULIRGs \citep[$<L_{6.2}/L_{3.3}> = 5.6$,][]{imanishi06a,imanishi07}.  Here we have used only isolated and compact galaxies in \citet{imanishi06a,imanishi07}, to minimize the effects due to different slit widths between the L-band and Spitzer slits, though \citet{imanishi08} compare their L-band spectra with Akari slitless spectra and find no evidence for significant slit loss.  The $L_{3.3}/L_{IR} \sim 8.5\pm4.7\times10^{-4}$ ratio (or $L_{3.3}/L_{FIR} \sim 1.5\times10^{-3}$ when using far-IR rather than total IR luminosity) is consistent with the ratio measured in local starburst galaxies \citep[$L_{3.3}/L_{FIR}\sim 1\times10^{-3}$][]{mouri90,imanishi02}, suggesting that this PAH feature may be used as an indicator of $L_{IR}$ at high redshift.  We note that the measured $L_{3.3}/L_{IR}$  is about five times lower than the ratio inferred from a recent broadband ($Spitzer$ IRAC) estimate of the 3.3 $\mu$m feature in LIRGs at $z\sim0.7$ \citep{magnelli08}.  

\section{Conclusions}

Due to the strong magnification (factor of $\sim28\times$) of the Cosmic Eye, we are able to perform a $Spitzer$ infrared study of an otherwise inaccessible $L^*_{UV}$ LBG.  We obtain high $S/N$ ($>5\sigma$) detections at 16, 24, and 70 $\mu$m, as well as high $S/N$ IRS Short-Low and Long-Low spectra from 7.4--14.5 $\mu$m and 19.5-35 $\mu$m, respectively.  With these data we compare different star formation diagnostics, and compare the IR properites to other star-forming galaxies at low and high redshift.  We find the following:

\begin{itemize}

\item Using the [O {\sc ii}] and maximum 16 $\mu$m fluxes of the foreground lens, we argue that far-IR photometry and mid-IR spectroscopy of the Cosmic Eye are not significantly contaminated by the lens.

\item The IR photometry of the eye (including a 3.5 mm flux limit) is fit by a relatively warm SED template and gives an $L_{IR}=8.3^{+4.7}_{-4.4}\times10^{11}$ L$_{\odot}$ after correction for magnification.  All of the IR SED templates give $L_{IR}$ estimates less than $1.3\times10^{12}$ L$_{\odot}$.  

\item The IRS Long-Low spectra show strong PAH emission at 6.2 and 7.7 $\mu$m that dominate the mid-IR luminosity.  The equivalent widths are near the maximum observed in local starburst galaxies \citep{brandl06,desai07,imanishi07} as well as high redshift ULIRGs with strong PAH emission \citep{sajina07,pope08}.  There are only small differences in the PAH flux ratios and overall spectral shape of the Cosmic Eye and composite spectra of local starbursts and high redshift SMGs, as well as the lensed LBG, cB58.  However, the $L_{7.7}/L_{6.2}$ PAH ratio of the SMG composite is significantly higher than that of both lensed LBGs.  The Cosmic Eye lies along the $L_{PAH}$-to-$L_{IR}$ correlations seen in starburst galaxies spanning three orders of magnitude in luminosity \citep{pope08,rigby08,menendez-delmestre09}.  Confirming this correlation in LBGs is important, as much of the IR-derived SFRs are currently based on 24 $\mu$m fluxes alone, which are dominated by PAH emission.

\item In the IRS Short-Low spectrum we detect the 3.3 $\mu$m PAH.  This is the highest redshift detection of this line and only the second reported detection outside of the local universe \citep[$z>0.2$,][]{sajina09}.  The PAH ratio, $L_{6.2}/L_{3.3} = 5.8\pm1.7$ is similar to the average ratio observed ($\sim6$) in local ULIRGs \citep{imanishi06a,imanishi07} and the 3.3-to-$L_{IR}$ ratio, $L_{3.3}/L_{IR} = 8.5\times10^{-4}$, is consistent with measurements from local starbursts \citep{mouri90,imanishi02}.  This line is of particular interest as it is the only strong PAH feature accessible with JWST at $z>3$ and may greatly facilitate studies of dust in the early universe.  Further measurements (in other LBGs) of the 3.3 $\mu$m feature's relation to other PAHs and to $L_{IR}$ are needed in order to interpret JWST measurements in the future.

\item Given an intrinsic UV spectral slope and a reddening curve, the observed spectral slope should predict the amount of IR emission (reprocessed UV light absorbed by dust) relative to unabsorbed UV light.  The measured $L_{FIR}/L_{UV}$ ratio is eight times lower than predicted from the UV spectral slope when assuming a Calzetti reddening curve.  That is, the Eye lies significantly below the IRX-$\beta$ relation measured in local starbursts \citep{meurer99}.  This has also been observed in the only other LBG to have detailed $Spitzer$ data, cB58 \citep{siana08b}.  Steeper extinction curves such as the LMC or SMC curves can rectify this apparent discrepancy.  \citet{siana08b} argue that the extinction curve is steeper than a Calzetti curve because much of the dust is in an outflowing foreground sheet with a large covering factor (near unity).  The Cosmic Eye, as with cB58, exhibits opaque absorption lines from outflowing gas, indicative of a high covering fraction of outflowing dust.  Therefore, we argue that both of these LBGs have steeper extinction curves than the Calzetti law due to dust geometries that differ from the patchy extinction seen in local starbursts.  However, a different dust composition can not be ruled out as a possible explanation for the discrepant reddening law.  

\end{itemize}

\acknowledgments

IRS, KEKC and ACE acknowledge support from STFC.  AMS also acknowledges support from the RAS Lockyer Fellowship.  JR acknowledges support from an EU Marie Curie Fellowship.

This work is based on observations made with the {\it Spitzer Space Telescope},
which is operated by the Jet Propulsion Laboratory, California Institute of 
Technology under a contract with NASA.  Support for this work was provided by 
NASA through an award issued by JPL/Caltech.

{\it Facilities:} \facility{Spitzer}

\clearpage



\clearpage

\bibliographystyle{apj}
\bibliography{apj-jour,all_ref}

\begin{thebibliography}{69}
\expandafter\ifx\csname natexlab\endcsname\relax\def\natexlab#1{#1}\fi

\bibitem[{{Adelberger} \& {Steidel}(2000)}]{adelberger00}
{Adelberger}, K.~L., \& {Steidel}, C.~C. 2000, \apj, 544, 218

\bibitem[{{Allam} {et~al.}(2007){Allam}, {Tucker}, {Lin}, {Diehl}, {Annis},
  {Buckley-Geer}, \& {Frieman}}]{allam07}
{Allam}, S.~S., {Tucker}, D.~L., {Lin}, H., {Diehl}, H.~T., {Annis}, J.,
  {Buckley-Geer}, E.~J., \& {Frieman}, J.~A. 2007, \apjl, 662, L51

\bibitem[{{Bell}(2002)}]{bell02}
{Bell}, E.~F. 2002, \apj, 577, 150

\bibitem[{{Belokurov} {et~al.}(2007){Belokurov}, {Evans}, {Moiseev}, {King},
  {Hewett}, {Pettini}, {Wyrzykowski}, {McMahon}, {Smith}, {Gilmore}, {Sanchez},
  {Udalski}, {Koposov}, {Zucker}, \& {Walcher}}]{belokurov07}
{Belokurov}, V. {et~al.} 2007, \apjl, 671, L9

\bibitem[{{Bendo} {et~al.}(2006{\natexlab{a}}){Bendo}, {Buckalew}, {Dale},
  {Draine}, {Joseph}, {Kennicutt}, {Sheth}, {Smith}, {Walter}, {Calzetti},
  {Cannon}, {Engelbracht}, {Gordon}, {Helou}, {Hollenbach}, {Murphy}, \&
  {Roussel}}]{bendo06}
{Bendo}, G.~J. {et~al.} 2006{\natexlab{a}}, \apj, 645, 134

\bibitem[{{Bendo} {et~al.}(2006{\natexlab{b}}){Bendo}, {Dale}, {Draine},
  {Engelbracht}, {Kennicutt}, {Calzetti}, {Gordon}, {Helou}, {Hollenbach},
  {Li}, {Murphy}, {Prescott}, \& {Smith}}]{bendo06b}
---. 2006{\natexlab{b}}, \apj, 652, 283

\bibitem[{{Brandl} {et~al.}(2006){Brandl}, {Bernard-Salas}, {Spoon}, {Devost},
  {Sloan}, {Guilles}, {Wu}, {Houck}, {Weedman}, {Armus}, {Appleton}, {Soifer},
  {Charmandaris}, {Hao}, {Higdon}, \& {Herter}}]{brandl06}
{Brandl}, B.~R. {et~al.} 2006, \apj, 653, 1129

\bibitem[{{Bruzual}(2007)}]{bruzual07}
{Bruzual}, G. 2007, ArXiv Astrophysics e-prints

\bibitem[{{Calzetti}(1997)}]{calzetti97}
{Calzetti}, D. 1997, \aj, 113, 162

\bibitem[{{Calzetti} {et~al.}(2000){Calzetti}, {Armus}, {Bohlin}, {Kinney},
  {Koornneef}, \& {Storchi-Bergmann}}]{calzetti00}
{Calzetti}, D., {Armus}, L., {Bohlin}, R.~C., {Kinney}, A.~L., {Koornneef}, J.,
  \& {Storchi-Bergmann}, T. 2000, \apj, 533, 682

\bibitem[{{Calzetti} {et~al.}(2005){Calzetti}, {Kennicutt}, {Bianchi},
  {Thilker}, {Dale}, {Engelbracht}, {Leitherer}, {Meyer}, {Sosey}, {Mutchler},
  {Regan}, {Thornley}, {Armus}, {Bendo}, {Boissier}, {Boselli}, {Draine},
  {Gordon}, {Helou}, {Hollenbach}, {Kewley}, {Madore}, {Martin}, {Murphy},
  {Rieke}, {Rieke}, {Roussel}, {Sheth}, {Smith}, {Walter}, {White}, {Yi},
  {Scoville}, {Polletta}, \& {Lindler}}]{calzetti05}
{Calzetti}, D. {et~al.} 2005, \apj, 633, 871

\bibitem[{{Calzetti} {et~al.}(1994){Calzetti}, {Kinney}, \&
  {Storchi-Bergmann}}]{calzetti94}
{Calzetti}, D., {Kinney}, A.~L., \& {Storchi-Bergmann}, T. 1994, \apj, 429, 582

\bibitem[{{Carilli} {et~al.}(2008){Carilli}, {Lee}, {Capak}, {Schinnerer},
  {Lee}, {McCraken}, {Yun}, {Scoville}, {Smol{\v c}i{\'c}}, {Giavalisco},
  {Datta}, {Taniguchi}, \& {Urry}}]{carilli08}
{Carilli}, C.~L. {et~al.} 2008, \apj, 689, 883

\bibitem[{{Chapman} {et~al.}(2005){Chapman}, {Blain}, {Smail}, \&
  {Ivison}}]{chapman05}
{Chapman}, S.~C., {Blain}, A.~W., {Smail}, I., \& {Ivison}, R.~J. 2005, \apj,
  622, 772

\bibitem[{{Chapman} {et~al.}(2000){Chapman}, {Scott}, {Steidel}, {Borys},
  {Halpern}, {Morris}, {Adelberger}, {Dickinson}, {Giavalisco}, \&
  {Pettini}}]{chapman00}
{Chapman}, S.~C. {et~al.} 2000, \mnras, 319, 318

\bibitem[{{Chary} \& {Elbaz}(2001)}]{chary01}
{Chary}, R., \& {Elbaz}, D. 2001, \apj, 556, 562

\bibitem[{{Coppin} {et~al.}(2007){Coppin}, {Swinbank}, {Neri}, {Cox}, {Smail},
  {Ellis}, {Geach}, {Siana}, {Teplitz}, {Dye}, {Kneib}, {Edge}, \&
  {Richard}}]{coppin07}
{Coppin}, K.~E.~K. {et~al.} 2007, \apj, 665, 936

\bibitem[{{Desai} {et~al.}(2007){Desai}, {Armus}, {Spoon}, {Charmandaris},
  {Bernard-Salas}, {Brandl}, {Farrah}, {Soifer}, {Teplitz}, {Ogle}, {Devost},
  {Higdon}, {Marshall}, \& {Houck}}]{desai07}
{Desai}, V. {et~al.} 2007, \apj, 669, 810

\bibitem[{{Draine} \& {Li}(2007)}]{draine07a}
{Draine}, B.~T., \& {Li}, A. 2007, \apj, 657, 810

\bibitem[{{Dye} {et~al.}(2007){Dye}, {Smail}, {Swinbank}, {Ebeling}, \&
  {Edge}}]{dye07}
{Dye}, S., {Smail}, I., {Swinbank}, A.~M., {Ebeling}, H., \& {Edge}, A.~C.
  2007, \mnras, 379, 308

\bibitem[{{Ellingson} {et~al.}(1996){Ellingson}, {Yee}, {Bechtold}, \&
  {Elston}}]{ellingson96}
{Ellingson}, E., {Yee}, H.~K.~C., {Bechtold}, J., \& {Elston}, R. 1996, \apjl,
  466, L71+

\bibitem[{{Engelbracht} {et~al.}(2006){Engelbracht}, {Kundurthy}, {Gordon},
  {Rieke}, {Kennicutt}, {Smith}, {Regan}, {Makovoz}, {Sosey}, {Draine},
  {Helou}, {Armus}, {Calzetti}, {Meyer}, {Bendo}, {Walter}, {Hollenbach},
  {Cannon}, {Murphy}, {Dale}, {Buckalew}, \& {Sheth}}]{engelbracht06}
{Engelbracht}, C.~W. {et~al.} 2006, \apjl, 642, L127

\bibitem[{{Erb} {et~al.}(2006){Erb}, {Steidel}, {Shapley}, {Pettini}, {Reddy},
  \& {Adelberger}}]{erb06}
{Erb}, D.~K., {Steidel}, C.~C., {Shapley}, A.~E., {Pettini}, M., {Reddy},
  N.~A., \& {Adelberger}, K.~L. 2006, \apj, 647, 128

\bibitem[{{Fitzpatrick}(1986)}]{fitzpatrick86}
{Fitzpatrick}, E.~L. 1986, \aj, 92, 1068

\bibitem[{{Frayer} {et~al.}(2006){Frayer}, {Huynh}, {Chary}, {Dickinson},
  {Elbaz}, {Fadda}, {Surace}, {Teplitz}, {Yan}, \& {Mobasher}}]{frayer06}
{Frayer}, D.~T. {et~al.} 2006, \apjl, 647, L9

\bibitem[{{Gil de Paz} {et~al.}(2005){Gil de Paz}, {Madore}, {Boissier},
  {Swaters}, {Popescu}, {Tuffs}, {Sheth}, {Kennicutt}, {Bianchi}, {Thilker}, \&
  {Martin}}]{gildepaz05}
{Gil de Paz}, A. {et~al.} 2005, \apjl, 627, L29

\bibitem[{{Goldader} {et~al.}(2002){Goldader}, {Meurer}, {Heckman}, {Seibert},
  {Sanders}, {Calzetti}, \& {Steidel}}]{goldader02}
{Goldader}, J.~D., {Meurer}, G., {Heckman}, T.~M., {Seibert}, M., {Sanders},
  D.~B., {Calzetti}, D., \& {Steidel}, C.~C. 2002, \apj, 568, 651

\bibitem[{{Holwerda} {et~al.}(2009){Holwerda}, {Keel}, {Williams}, {Dalcanton},
  \& {de Jong}}]{holwerda09}
{Holwerda}, B.~W., {Keel}, W.~C., {Williams}, B., {Dalcanton}, J.~J., \& {de
  Jong}, R.~S. 2009, \aj, 137, 3000

\bibitem[{{Hummer} \& {Storey}(1987)}]{hummer87}
{Hummer}, D.~G., \& {Storey}, P.~J. 1987, \mnras, 224, 801

\bibitem[{{Imanishi}(2002)}]{imanishi02}
{Imanishi}, M. 2002, \apj, 569, 44

\bibitem[{{Imanishi} \& {Dudley}(2000)}]{imanishi00}
{Imanishi}, M., \& {Dudley}, C.~C. 2000, \apj, 545, 701

\bibitem[{{Imanishi} {et~al.}(2007){Imanishi}, {Dudley}, {Maiolino}, {Maloney},
  {Nakagawa}, \& {Risaliti}}]{imanishi07}
{Imanishi}, M., {Dudley}, C.~C., {Maiolino}, R., {Maloney}, P.~R., {Nakagawa},
  T., \& {Risaliti}, G. 2007, \apjs, 171, 72

\bibitem[{{Imanishi} {et~al.}(2006){Imanishi}, {Dudley}, \&
  {Maloney}}]{imanishi06a}
{Imanishi}, M., {Dudley}, C.~C., \& {Maloney}, P.~R. 2006, \apj, 637, 114

\bibitem[{{Imanishi} {et~al.}(2008){Imanishi}, {Nakagawa}, {Ohyama},
  {Shirahata}, {Wada}, {Onaka}, \& {Oi}}]{imanishi08}
{Imanishi}, M., {Nakagawa}, T., {Ohyama}, Y., {Shirahata}, M., {Wada}, T.,
  {Onaka}, T., \& {Oi}, N. 2008, \pasj, 60, 489

\bibitem[{{Kennicutt}(1998)}]{kennicutt98}
{Kennicutt}, Jr., R.~C. 1998, \araa, 36, 189

\bibitem[{{Latter}(1991)}]{latter91}
{Latter}, W.~B. 1991, \apj, 377, 187

\bibitem[{{Lin} {et~al.}(2008){Lin}, {Buckley-Geer}, {Allam}, {Tucker},
  {Diehl}, {Kubik}, {Kubo}, {Annis}, {Frieman}, {Oguri}, \& {Inada}}]{lin09}
{Lin}, H. {et~al.} 2008, ArXiv e-prints

\bibitem[{{Magnelli} {et~al.}(2008){Magnelli}, {Chary}, {Pope}, {Elbaz},
  {Morrison}, \& {Dickinson}}]{magnelli08}
{Magnelli}, B., {Chary}, R.~R., {Pope}, A., {Elbaz}, D., {Morrison}, G., \&
  {Dickinson}, M. 2008, \apj, 681, 258

\bibitem[{{Maiolino} {et~al.}(2004){Maiolino}, {Schneider}, {Oliva}, {Bianchi},
  {Ferrara}, {Mannucci}, {Pedani}, \& {Roca Sogorb}}]{maiolino04}
{Maiolino}, R., {Schneider}, R., {Oliva}, E., {Bianchi}, S., {Ferrara}, A.,
  {Mannucci}, F., {Pedani}, M., \& {Roca Sogorb}, M. 2004, \nat, 431, 533

\bibitem[{{Makovoz} \& {Khan}(2005)}]{makovoz05b}
{Makovoz}, D., \& {Khan}, I. 2005, in Astronomical Society of the Pacific
  Conference Series, Vol. 347, Astronomical Data Analysis Software and Systems
  XIV, ed. P.~{Shopbell}, M.~{Britton}, \& R.~{Ebert}, 81--+

\bibitem[{{Makovoz} \& {Marleau}(2005)}]{makovoz05a}
{Makovoz}, D., \& {Marleau}, F.~R. 2005, \pasp, 117, 1113

\bibitem[{{Men{\'e}ndez-Delmestre} {et~al.}(2009){Men{\'e}ndez-Delmestre},
  {Blain}, {Smail}, {Alexander}, {Chapman}, {Armus}, {Frayer}, {Ivison}, \&
  {Teplitz}}]{menendez-delmestre09}
{Men{\'e}ndez-Delmestre}, K. {et~al.} 2009, ArXiv e-prints

\bibitem[{{Meurer} {et~al.}(1999){Meurer}, {Heckman}, \& {Calzetti}}]{meurer99}
{Meurer}, G.~R., {Heckman}, T.~M., \& {Calzetti}, D. 1999, \apj, 521, 64

\bibitem[{{Moorwood}(1986)}]{moorwood86}
{Moorwood}, A.~F.~M. 1986, \aap, 166, 4

\bibitem[{{Mouri} {et~al.}(1990){Mouri}, {Kawara}, {Taniguchi}, \&
  {Nishida}}]{mouri90}
{Mouri}, H., {Kawara}, K., {Taniguchi}, Y., \& {Nishida}, M. 1990, \apjl, 356,
  L39

\bibitem[{{Papovich} {et~al.}(2001){Papovich}, {Dickinson}, \&
  {Ferguson}}]{papovich01}
{Papovich}, C., {Dickinson}, M., \& {Ferguson}, H.~C. 2001, \apj, 559, 620

\bibitem[{{Pettini} \& {Pagel}(2004)}]{pettini04}
{Pettini}, M., \& {Pagel}, B.~E.~J. 2004, \mnras, 348, L59

\bibitem[{{Pettini} {et~al.}(2000){Pettini}, {Steidel}, {Adelberger},
  {Dickinson}, \& {Giavalisco}}]{pettini00}
{Pettini}, M., {Steidel}, C.~C., {Adelberger}, K.~L., {Dickinson}, M., \&
  {Giavalisco}, M. 2000, \apj, 528, 96

\bibitem[{{Pilyugin} \& {Thuan}(2005)}]{pilyugin05}
{Pilyugin}, L.~S., \& {Thuan}, T.~X. 2005, \apj, 631, 231

\bibitem[{{Pope} {et~al.}(2008){Pope}, {Chary}, {Alexander}, {Armus},
  {Dickinson}, {Elbaz}, {Frayer}, {Scott}, \& {Teplitz}}]{pope08}
{Pope}, A. {et~al.} 2008, \apj, 675, 1171

\bibitem[{{Prevot} {et~al.}(1984){Prevot}, {Lequeux}, {Prevot}, {Maurice}, \&
  {Rocca-Volmerange}}]{prevot84}
{Prevot}, M.~L., {Lequeux}, J., {Prevot}, L., {Maurice}, E., \&
  {Rocca-Volmerange}, B. 1984, \aap, 132, 389

\bibitem[{{Reddy} \& {Steidel}(2004)}]{reddy04}
{Reddy}, N.~A., \& {Steidel}, C.~C. 2004, \apjl, 603, L13

\bibitem[{{Reddy} \& {Steidel}(2009)}]{reddy09}
---. 2009, \apj, 692, 778

\bibitem[{{Reddy} {et~al.}(2006){Reddy}, {Steidel}, {Fadda}, {Yan}, {Pettini},
  {Shapley}, {Erb}, \& {Adelberger}}]{reddy06}
{Reddy}, N.~A., {Steidel}, C.~C., {Fadda}, D., {Yan}, L., {Pettini}, M.,
  {Shapley}, A.~E., {Erb}, D.~K., \& {Adelberger}, K.~L. 2006, \apj, 644, 792

\bibitem[{{Rigby} {et~al.}(2008){Rigby}, {Marcillac}, {Egami}, {Rieke},
  {Richard}, {Kneib}, {Fadda}, {Willmer}, {Borys}, {van der Werf},
  {P{\'e}rez-Gonz{\'a}lez}, {Knudsen}, \& {Papovich}}]{rigby08}
{Rigby}, J.~R. {et~al.} 2008, \apj, 675, 262

\bibitem[{{Risaliti} {et~al.}(2006){Risaliti}, {Maiolino}, {Marconi}, {Sani},
  {Berta}, {Braito}, {Ceca}, {Franceschini}, \& {Salvati}}]{risaliti06}
{Risaliti}, G. {et~al.} 2006, \mnras, 365, 303

\bibitem[{{Sajina} {et~al.}(2007){Sajina}, {Yan}, {Armus}, {Choi}, {Fadda},
  {Helou}, \& {Spoon}}]{sajina07}
{Sajina}, A., {Yan}, L., {Armus}, L., {Choi}, P., {Fadda}, D., {Helou}, G., \&
  {Spoon}, H. 2007, \apj, 664, 713

\bibitem[{{Sajina} {et~al.}(2009){Sajina}, {Yan}, {Spoon}, \&
  {Fadda}}]{sajina09}
{Sajina}, A., {Yan}, L., {Spoon}, H., \& {Fadda}, D. 2009, ArXiv e-prints

\bibitem[{{Shapley} {et~al.}(2001){Shapley}, {Steidel}, {Adelberger},
  {Dickinson}, {Giavalisco}, \& {Pettini}}]{shapley01}
{Shapley}, A.~E., {Steidel}, C.~C., {Adelberger}, K.~L., {Dickinson}, M.,
  {Giavalisco}, M., \& {Pettini}, M. 2001, \apj, 562, 95

\bibitem[{{Siana} {et~al.}(2008){Siana}, {Teplitz}, {Chary}, {Colbert}, \&
  {Frayer}}]{siana08b}
{Siana}, B., {Teplitz}, H.~I., {Chary}, R.-R., {Colbert}, J., \& {Frayer},
  D.~T. 2008, \apj, 689, 59

\bibitem[{{Smail} {et~al.}(2007){Smail}, {Swinbank}, {Richard}, {Ebeling},
  {Kneib}, {Edge}, {Stark}, {Ellis}, {Dye}, {Smith}, \& {Mullis}}]{smail07}
{Smail}, I. {et~al.} 2007, \apjl, 654, L33

\bibitem[{{Stark} {et~al.}(2008){Stark}, {Swinbank}, {Ellis}, {Dye}, {Smail},
  \& {Richard}}]{stark08}
{Stark}, D.~P., {Swinbank}, A.~M., {Ellis}, R.~S., {Dye}, S., {Smail}, I.~R.,
  \& {Richard}, J. 2008, \nat, 455, 775

\bibitem[{{Steidel} {et~al.}(1996){Steidel}, {Giavalisco}, {Pettini},
  {Dickinson}, \& {Adelberger}}]{steidel96}
{Steidel}, C.~C., {Giavalisco}, M., {Pettini}, M., {Dickinson}, M., \&
  {Adelberger}, K.~L. 1996, \apjl, 462, L17+

\bibitem[{{Stratta} {et~al.}(2007){Stratta}, {Maiolino}, {Fiore}, \&
  {D'Elia}}]{stratta07}
{Stratta}, G., {Maiolino}, R., {Fiore}, F., \& {D'Elia}, V. 2007, \apjl, 661,
  L9

\bibitem[{{Teplitz} {et~al.}(2007){Teplitz}, {Desai}, {Armus}, {Chary},
  {Marshall}, {Colbert}, {Frayer}, {Pope}, {Blain}, {Spoon}, {Charmandaris}, \&
  {Scott}}]{teplitz07}
{Teplitz}, H.~I. {et~al.} 2007, \apj, 659, 941

\bibitem[{{Teplitz} {et~al.}(2000){Teplitz}, {McLean}, {Becklin}, {Figer},
  {Gilbert}, {Graham}, {Larkin}, {Levenson}, \& {Wilcox}}]{teplitz00}
---. 2000, \apjl, 533, L65

\bibitem[{{Thilker} {et~al.}(2005){Thilker}, {Bianchi}, {Boissier}, {Gil de
  Paz}, {Madore}, {Martin}, {Meurer}, {Neff}, {Rich}, {Schiminovich},
  {Seibert}, {Wyder}, {Barlow}, {Byun}, {Donas}, {Forster}, {Friedman},
  {Heckman}, {Jelinsky}, {Lee}, {Malina}, {Milliard}, {Morrissey}, {Siegmund},
  {Small}, {Szalay}, \& {Welsh}}]{thilker05}
{Thilker}, D.~A. {et~al.} 2005, \apjl, 619, L79

\bibitem[{{Tielens}(2008)}]{tielens08}
{Tielens}, A.~G.~G.~M. 2008, \araa, 46, 289

\bibitem[{{Todini} \& {Ferrara}(2001)}]{todini01}
{Todini}, P., \& {Ferrara}, A. 2001, \mnras, 325, 726

\end{thebibliography}

\end{document}